\documentclass[twocolumn]{aastex63}
\usepackage[version=4]{mhchem}
\usepackage{booktabs} 
\usepackage{graphics, multirow}
\usepackage{dblfloatfix}

\newcommand{\cmmnt}[1]{}
\usepackage{hyperref} 
\usepackage{graphics, multirow}
\usepackage{enumitem}
\usepackage{refcount}
\usepackage{float}
\usepackage[T1]{fontenc}

\usepackage{amsmath}	

\makeatletter
\let\scshape\relax 
\DeclareRobustCommand\scshape{%
  \not@math@alphabet\scshape\relax
  \ifnum\pdf@strcmp{\f@family}{\familydefault}=\z@
    \fontfamily{qbk}%
  \fi
  \fontshape\scdefault\selectfont}
\makeatother

\graphicspath{ {./figures/} }

\shorttitle{Signatures of Lithium-bearing Molecules in Brown Dwarfs}
\shortauthors{Gharib-Nezhad $\&$ Marley et al.}

\begin{document}

\title{Following the Lithium: Tracing Li-bearing Molecules Across Age, Mass, and Gravity in Brown Dwarfs}

\author[0000-0002-4088-7262]{Ehsan Gharib-Nezhad}\thanks{NASA Postdoctoral Fellow\\corresponding author: \href{mailto:e.gharibnezhad@asu.edu}{e.gharibnezhad@asu.edu}}
\affiliation{NASA Ames Research Center, Moffett Field, CA 94035, USA}

\author[0000-0002-5251-2943]{Mark S. Marley}
\affiliation{Lunar and Planetary Laboratory, University of Arizona, Tucson, Arizona 85721, USA}

\author[0000-0003-1240-6844]{Natasha E. Batalha}
\affiliation{NASA Ames Research Center, Moffett Field, CA 94035, USA}

\author[0000-0001-6627-6067]{Channon Visscher}
\affiliation{Chemistry $\&$ Planetary Sciences, Dordt University, Sioux Center, IA. 51250, USA}
\affiliation{Center for Extrasolar Planetary Systems, Space Science Institute, Boulder, CO 80301, USA}

\author{Richard S. Freedman}
\affiliation{SETI Institute, Mountain View, CA 94035, USA}
\affiliation{NASA Ames Research Center, Moffett Field, CA 94035, USA}

\author{Roxana E. Lupu}
\affiliation{BAER Institute/NASA Ames Research Center, Moffett Field, CA 94035, USA}


\begin{abstract}
{Lithium is an important element for the understanding of ultracool dwarfs because it is lost to fusion at masses above $\sim 68\, M_{\rm J}$. Hence, the presence or absence of atomic Li has served as an indicator of the nearby H-burning boundary at about $75\,M_{\rm J}$ between brown-dwarfs and very low-mass stars. Historically the ``Lithium test'', a search for the presence and strength of the Li line at 670.8 nm, has been a marker if an object has a substellar mass with stellar-like spectral energy distribution (e.g., a late-type M dwarf). While the Li test could in principle also be used to distinguish masses of later-type L-T dwarfs,  Li is predominantly no longer found as an atomic gas, but rather a molecular species such as LiH, LiF, LiOH, and LiCl in their cooler atmospheres. L- and T-type brown dwarfs are also quite faint at 670 nm and thus challenging targets for high resolution spectroscopy. But only recently have experimental molecular line lists become available for the molecular Li species, allowing molecular Li mass discrimination. In this study, we generated the latest opacity of each of these Li-bearing molecules and performed thermochemical equilibrium atmospheric composition calculation of the abundance of these molecules. Finally, we computed thermal emission spectra for a series of radiative-convective equilibrium models of cloudy and cloudless brown dwarf atmospheres (with $T_{\rm eff}=$ 500--2400~K, and $\log g$=4.0, 4.5, 5.0) to understand where the presence or absence of atmospheric lithium-bearing species is most easily detected as a function of brown dwarf mass and age. After atomic Li, the best spectral signatures were found to be LiF at $10.5-12.5\, \micron$ and LiCl at $14.5-18.5\, \micron$. LiH also shows a narrow feature at $\sim 9.38 \, \micron$. }
\end{abstract}

\keywords{}
\pagebreak
\clearpage
\section{Introduction}\label{sec:intro}

More than 25 years ago, the observation of the first brown dwarfs, { Teide 1} and Gliese 229B, ushered a new era in astronomy \citep{Rebolo1995Nature,Oppenheimer1995,Nakajima1995}. Since then, many models and observations have been done to understand the chemistry and physics of brown dwarf atmospheres, their atmospheric evolution with age, mass, and effective temperature \citep[e.g.,][]{Saumon2008L-T-dwarfs,Chabrier2000Evolution,Burrows1997EvolutionAtmo}, as well as their connection to the planet and stellar main characteristics \citep[e.g.,][]{Marley2015, Zhang2020Review}. Luminosity, age, mass, radius, surface gravity, atmospheric evolution, effective temperature, and cloud/aerosol formation  are among the main active open questions waiting to be addressed to better understand these objects.

One indicator used to distinguish between stars and brown dwarfs is the detection of the atomic lithium resonance line at 670.8~nm, which was proposed by  \citet{Rebolo1992-Li-Test} and aptly named the ``lithium test’’. Atomic Li is converted to He through fusion at central temperatures of $\sim2-3\times 10^{6}\,{\rm K}$. { Thus objects which are not massive enough to reach such temperatures in their interior do not burn Li.} \citet{Nelson1993} and \citet{Pozio1991} calculated a transition mass of $62-65\,M_{\rm J}$   for the Li burning. { \citet{Magazzu1993} showed 
that the required mass to burn Li by half is 84 and  $M=63\ M_{\rm J}$ for 60 Myr and  250 Myr objects, respectively.} Similarly,  \citet{Chabrier1996LiTest}  modeled the depletion of Li along with mass and age, and found that the initial lithium burns by a factor of 2 over $\sim$0.26~Gyr and by a factor of 100 over $\sim 1\,{\rm Gyr}$ for 64~$M_\mathrm{J}$ object. {  In addition, \citet{Bildsten1997} provided an analytical formations to calculate the Li depletion as a function of age, mass, radius, and luminosity of objects within 52--105$M_{\rm J}$.}
By comparison, current calculations place the mass limit for H burning at $74\,M_{\mathrm{J}}$ \citep[e.g.,][submitted]{Marley2021}. Because of the happy near  coincidence of these two masses, the detection of Li has become a straightforward diagnostic test of most brown dwarfs; even though, consumption of Li in the most massive brown dwarfs leads to their confusion with the  lowest mass stars \citep{Basri2000_Observ-BD_LiTest}. 

Interestingly \citet[][]{Rebolo96} further validated Teide 1 as a true brown dwarf through the detection of atomic Lithium and application of the Lithium test. 
Since then, several studies have implemented the Li test to confirm the true detection of brown dwarfs and also determine their ages and masses  \citep[e.g.,][]{Martin1999Litest-BD}. Age is difficult to ascertain, however, and systematic studies of the Li strength of a sample of $\sim$150 late-M to L dwarfs by \citet{Kirkpatrick2008-LiTest-Ldwarfs} did not reveal any strong correlation with age.  

In general, the Li test has been challenging to apply to all types of brown dwarfs. First, moderately large telescopes are required with a resolution $\lambda/\Delta\lambda \geqslant 1200$ and signal-to-noise ratio (SNR) of $\sim 50$ to detect the Li resonance line at 670.8~nm \citep{Kirkpatrick2008-LiTest-Ldwarfs,Martin1999Litest-BD}\footnote{  Note that lower SNR and resolution is required to record the Li resonance line in L dwarfs. For example, \citep{Martin2018} showed that SNR$>$15 and $\lambda/\Delta\lambda \geqslant 600$ are sufficient to detect the Li 6708 \AA absorption in L-type Hyades BD candidates}. Second, for the mass interval between this minimum Li-burning mass and the hydrogen-burning minimum mass $\sim 74\ M_{\rm J}$, absent atomic Lithium is ambiguous as a star/brown dwarf discriminant. This is particularly troublesome in the effective temperature range occupied by both young massive brown dwarfs and the coolest, lowest mass stars.

For objects cooler than any star at the bottom of the main sequence there is no question that an object is substellar (e.g., the atmospheric methane detected in Gl 229B, clearly indicated a substellar atmosphere \citep{Oppenheimer1995}). However, the Li test still has relevance even at these cool effective temperatures. Gravity signatures for brown dwarfs are currently lacking and beyond general rules-of-thumb such as the triangular-shaped H band as a signpost of low gravity in L dwarfs \citep{Kirkpatrick2008-LiTest-Ldwarfs}, there are no straightforward markers of mass in brown dwarfs.
As a result, many L and T-type dwarfs have uncertain masses. This is primarily due to the well known degeneracy between age, mass, and gravity.

{ Specific examples of objects with divergent mass estimates include Gliese 229 B  (late-T, $\sim$1000~K) for which  \citet{Brandt2020}, by combining radial velocities, imaging, and astrometry methods,  derived a dynamical mass of $\sim~70\pm 5 M_{\rm J}$, whereas early evolutionary models suggested a younger age and noticeably lower  mass of $20-55 M_\mathrm{J}$ \citep{Marley1996, Allard1996}. HD~19467B is another T~dwarf (T5-T7, $\sim$1050~K)  with high uncertainty in its modeled mass and age, which are $74^{+12}_{-9} M_\mathrm{J}$ and $\sim9-11$~Gyr \citep{Maire2020}. Another T-dwarf example is GJ~570D (T7.5, $\sim$800~K) with a calculated mass of 15--72~$M_{\mathrm{J}}$ \citep{Geballe2001}.  This degeneracy problem is not only limited to T dwarfs. HD~7672B and HD~4747B are two examples of massive L dwarfs (L4.5 and L8, $\sim$1700~K) with large uncertainties in their mass and age \citep{Liu2002, Crepp2016}. For objects such as these, a discrete mass discriminant would be particularly valuable. 

Note that the computed thermal evolution tracks of brown dwarfs \citep[e.g.,][]{Marley2021} predict that objects massive enough to burn Li would not cool below 900 K in the age of the universe. However as some of the above mass estimates attest, a few objects, like Gl 229 B, have reported apparently problematic masses. Having an independent mass indicator would greatly illuminate such cases.}


To the best of our knowledge, there has been only one study before by \citet{Weck2004} to investigate the  LiCl signature for dwarfs with $T_{\rm eff}$=900--1500~K. However, the main goals of their study were to provide the LiCl $ab\,initio$ line list and consider the brown dwarf effective temperature range over which LiCl could be detected. In comparison, we here assess a broader range of atmospheric temperatures and include all Li species for which there is opacity data available.

We have organized this article as follows; In \S~\ref{sec:LiChemistry}, we give an account of the equilibrium chemistry in M to T dwarfs, and the chemistry of the dominant Li-bearing molecules is explained. In  \S~\ref{sec:AbsCrosSec}, we describe the calculation of absorption cross-section (or opacity) data for LiH, LiF, and LiCl molecules. The degeneracy in the determination of brown dwarf masses is addressed in \S\ref{sec:degeneracy}.  Following that, \S~\ref{sec:modelingflux} presents a set of synthetic thermal fluxes for brown dwarfs with effective temperature in a range of 500--2400 K, and for surface gravity ($\log g$) of 4.0, 4.5 and  5.0. In this section, the presence and spectral location of LiH, LiF, and LiCl molecules in the cloudy and clear atmospheres is discussed. In the end, \S~\ref{sec:atomic-Li-line} comments on the non-detection of atomic Li in T dwarf observed fluxes even if modeling spectra predict its presence. Summary and future works are presented in \S~\ref{sec:Conclusion}.

\section{Lithium Chemistry} \label{sec:LiChemistry}

Thermochemical equilibrium simulations are performed to calculate the abundance of the following species in the temperature-pressure space at 75--6000~K and 10$^{-6}$--3000~bar: \ce{Li(g)}, \ce{LiOH}, \ce{LiH}, \ce{LiF}, \ce{LiCl}, \ce{e-},  \ce{H2}, \ce{H}, \ce{H+}, \ce{H-}, \ce{H2-}, \ce{H2+}, \ce{H3+}, \ce{H2O}, \ce{CH4}, \ce{C2H2}, \ce{C2H4}, \ce{C2H6}, \ce{CO}, \ce{CO2}, \ce{NH3}, \ce{N2}, \ce{HCN},  \ce{OCS}, \ce{PH3}, \ce{H2S}, \ce{TiO}, \ce{VO}, \ce{FeH}, \ce{CrH}, \ce{SiO}, \ce{MgH}, \ce{He}, \ce{Na}, \ce{K}, \ce{Rb}, \ce{Cs}, \ce{Fe}, \ce{graphite}, \ce{Li(s)}. Since we are exploring objects in which lithium is not expected to burn, we used the protosolar abundance of Li is used in this study. We adopt the protosolar Li abundance from \citet{Lodders2003Abundance} (i.e., A(Li)=3.28), which is derived from  meteoritic CI chondrites (carbonaceous meteorites of Ivuna type). CI chondrites are the most primitive chondrites because they have not chemically fractionated and their relative abundances are considered to be similar to protosolar abundance. Therefore, they are enriched in elements like lithium. 

\begin{figure*}[htb!]
\centering
\includegraphics[scale=0.7]{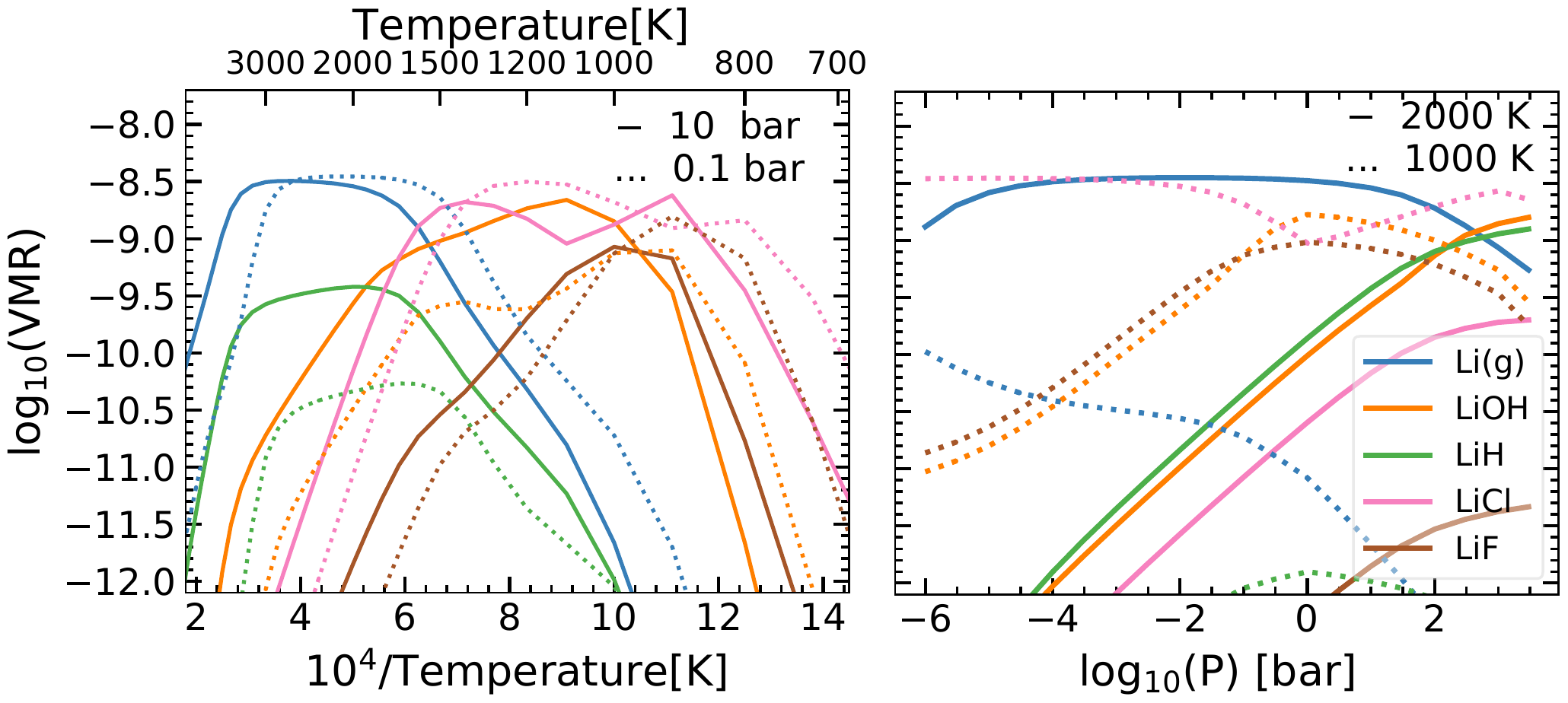} 
	  \caption{Volume Mixing ratio (VMR) plots of Li(g), LiF, LiCl, LiH, and LiOH with respect to temperature (left) and pressure (right). At high temperatures ($T\gtrsim 1500\,\rm K$), gaseous Li is the most abundant Li-bearing species. At moderate temperatures ($1500 \, {\rm K}\lesssim T \lesssim 1000\,\rm K$), both LiCl and LiOH will be the most stable form of Li-bearing species, depending on the pressure. At low pressures and low temperatures ($P\lesssim 0.01\,\rm bar$ and $T\lesssim 1000\,\rm K$), LiF could also be abundant.}
	  \label{fig:VMR} 
\end{figure*}

Figure~\ref{fig:VMR} illustrates the Volume Mixing Ratios (hereafter, VMR) of Li-bearing species predicted by equilibrium chemistry as a function of temperature (left) and pressure (right). There are three main thermal regions in the VMR-$T$-$P$ parameter space as defined by these species. First, hot atmospheres with $T_{\rm eff}$ in $\sim$1800--2800~K (e.g., M-L type dwarfs) in which monoatomic Li is the dominant gas in $\sim 10^{-3}$--100~bar. Second, warm atmospheres with $T_{\rm eff}$ in $\sim$900-1800~K (e.g., L-T type dwarfs) where LiCl and LiOH dominate. The third region, with T$_{\rm eff}\leqslant$1000, is characterized by LiF and LiCl as the favored molecules.

\begin{figure*}[htb!]
\centering
\includegraphics[scale=0.7]{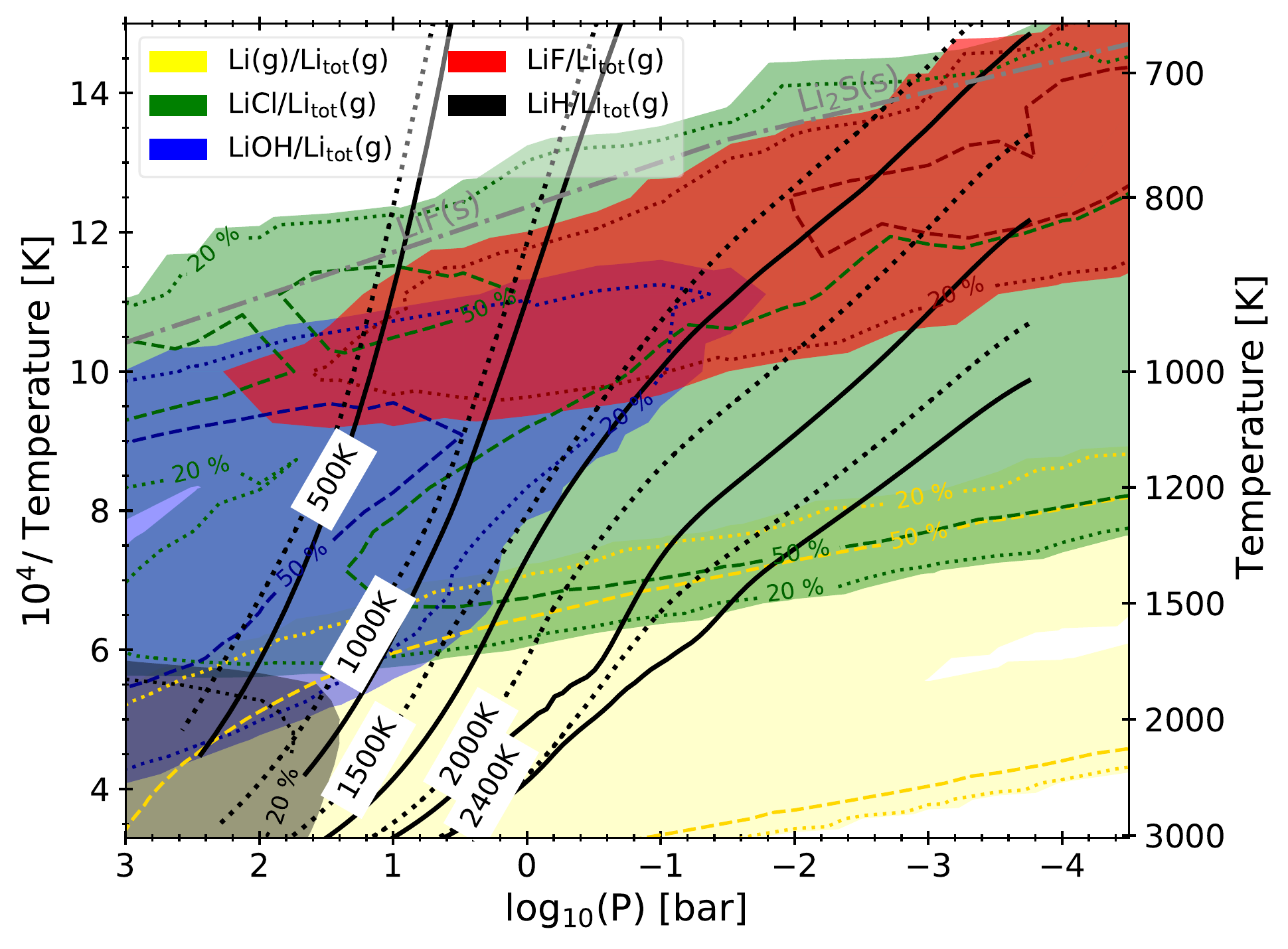} 
	  \caption{Li phase diagram: temperature-pressure diagram showing the mole fraction of atomic gaseous Li (yellow), LiH (black), LiOH (blue), LiCl (green), and LiF (red). For each area, the dotted and dashed contours represent the 20\% and 50\% of the species with respect to the total Lithium abundance in both solid and gas phases (i.e., the protosolar value $\sim3.5\times10^{-9}$). The solid and dashed black curves are the  $T(P)$ profiles (for $\log g$=5) for cloudy and cloudless \citep{Marley2021} models, respectively. As a brown dwarf cools, the dominant Li-bearing species in the observable atmosphere (generally $P<10\,\rm bar$) varies.}
	  \label{fig:Li-PhaseDiagram} 
\end{figure*}

Figure~\ref{fig:Li-PhaseDiagram} illustrates a broad view of the presence and abundance of Li species in the temperature-pressure space. In this phase diagram, Li, LiCl, LiOH, and LiF are shown by yellow, green, blue, and red shading. For example, gaseous Li is the most stable form at  $T>1400$~K. At each area, dashed and dotted contour lines represent 50$\%$ and 20$\%$ of that species with respect to the total protosolar lithium abundance  ($\sim3.5\times10^{-9}$) in the atmosphere. The solid and dashed black curves are the  $T(P)$ profiles (for $\log g$=5) for cloudy and cloudless \citep{Marley2021} models, respectively. We aim to understand  whether these  different Li species domains can be recognized in brown dwarf spectra and if they could serve as mass indicators.

\section{Absorption Cross-Sections}\label{sec:AbsCrosSec}
Radiative transfer modeling of synthetic thermal emission requires accurate and complete absorption cross-section data (hereafter, ACS) of Li (g), LiH, LiF, and LiCl. ACS data of a given absorber consists of several million lines with unique position, intensity, and width. The width is sensitive to the temperature and pressure of broadening gases, and it is controlled by the convolution of Doppler and Lorentz profiles. The knowledge of pressure broadening coefficients for the Lorentz profile is instrumental and is dependent on $J$ quantum numbers. The choice-of-broadening for the background gases in this study is 85$\%$ \ce{H2} and 15$\%$ He by number.  Because the \ce{H2} and \ce{He} pressure-broadening Lorentz coefficients for these species have not been calculated or measured, we have estimated pressure-broadening data for these Li-broadening molecules in question. 

In our recent paper by \citet{Gharib-Nezhad2021}, we provided a detailed discussion on how to generate these ACS data and their pressure-broadening coefficients. The generated ACS data of atomic Li(g), LiH, LiF, and LiCl covers 1460 temperature and pressure grid points, and range of 75--4000~K and 10$^{-6}$--3000~bar, respectively. The line lists used to generate these ACS data are the most accurate and up-to-date, which are provided in Table~\ref{tab:Summary-linelist}. Note that the LiOH line list is not available, and hence we could not generate the LiOH ACS data.          

\begin{figure*}[htb!]
\centering
\includegraphics[scale=.6]{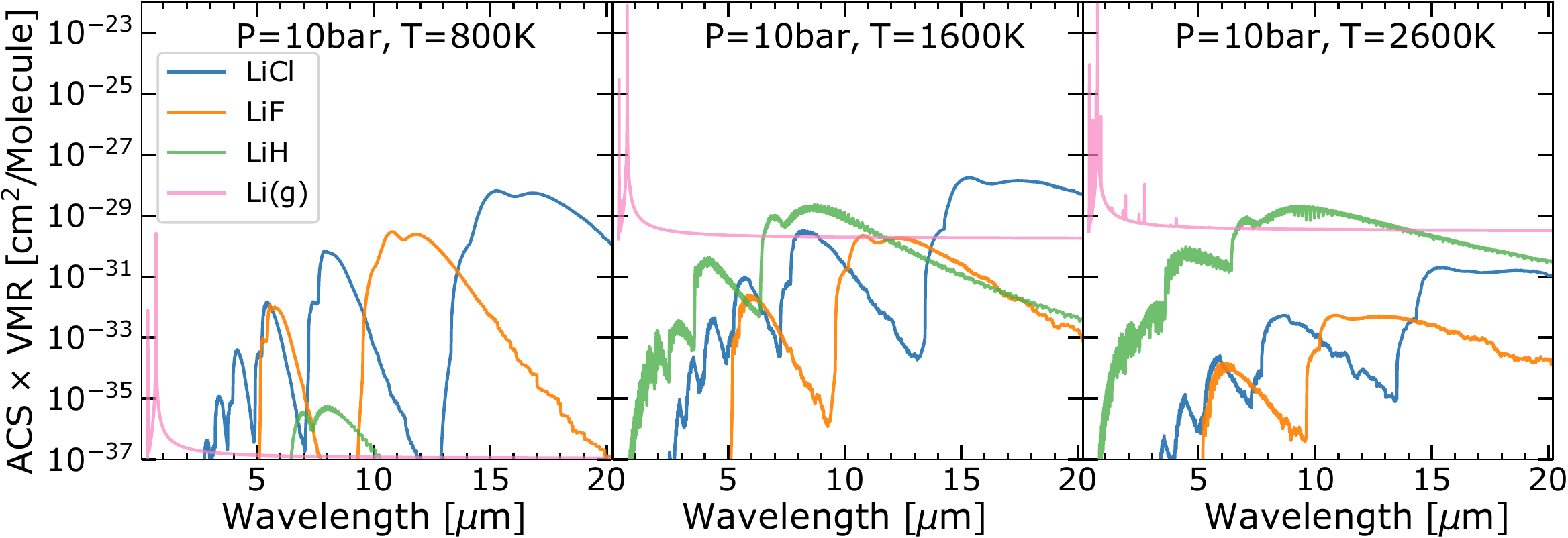} 
	  \caption{Smoothed absorption cross-sections (ACSs) of atomic Li, LiH, LiF, and LiCl multiplied by its volume mixing ratios (VMR) for $P=10$~bar. The resonance Li line at 670.8~nm has a large ACS value than its other molecular forms, and so its ACS$\times$VMR value is larger than others even at low temperatures. In contrast, LiCl and LiF are the two dominant molecules at low temperatures and their spectral features are clearly distinguishable. At $1600\,\rm K$, the fundamental band of LiH at $\sim 8.5\, \rm \micron$ is dominant. In this study, their latest line lists are used to compute their opacities and also their pressure-broadening coefficients are revised. See Table~\ref{tab:Summary-linelist} for more details. }
	  \label{fig:LiX-XS} 
\end{figure*}

Figure~\ref{fig:LiX-XS} depicts the weighted ACS$\times$VMR for Li, LiH, LiF, and LiCl for 800, 1600, and 2600~K at $P=$10~bar. The dominant spectral features LiH, LiF, and LiCl are approximately located at $\sim$9 -- 10 $\micron$, 10.5 -- 14 $\micron$, and 14.5 -- 19 $\micron$,  respectively. Although the atomic Li is not stable at $T<1000$~K, its ACS value is a few order-of-magnitude greater than its molecular forms. Hence, the Li absorption is predominant in the optical region and, as we will show, is present even after molecular species become predominant in cooler atmospheric layers. At $T<$2000~K, LiCl and LiF become predominant features in the infrared.     

\begin{table}[htp!]
  \footnotesize 
  \caption{Summary list of opacities: molecules, temperature, pressures, and their line lists}\label{tab:Summary-linelist}   
\centering 
\begin{tabular}{ l l l l } 
\hline\hline
Absorber	 &	$\lambda$ [$\mu$m]  & Method & Reference\\
\hline
LiCl  	   &     2.2--330            & laboratory      &   \citep{Bittner2018} \\
\hline
LiF  	   &     5.5--330            & laboratory      &   \citep{Bittner2018} \\
\hline
LiH  	   &     5.5--330            & ab initio     &   \citep{Coppola2011} \\
\hline
\end{tabular}
\centering 
       \scriptsize
       \hfill\parbox[t]{\linewidth}{
       Note: We computed the ACS over the pressure and temperature ranges of 10$^{-6}$--3000~bar and 75-4000~K. See \S\ref{sec:AbsCrosSec} for more details.}

\end{table}

\section{Degeneracy on age-mass-gravity space: Challenges}\label{sec:degeneracy}

Figure \ref{fig:g_T_t_M} presents in black the Sonora-Bobcat solar-metallicity evolution tracks for isolated 10 to 83~$M_{\rm J}$ dwarfs in $\log g-T_{\rm eff}$ parameter space from \citet{Marley2021}. Green curves show isochrones from 0.04 to 10~Gyr. This figure also shows the Li depletion region (yellow) where atomic Li is depleted by fusion. The presence and overlap of LiF, LiCl, and LiH molecules are highlighted as well. 

Our understanding of brown dwarfs is shaped by the degeneracy in the $T_{\rm eff}-t-g-M-R-L$ space (i.e., effective temperature, age, surface gravity, mass, radius, luminosity). The $T_{\rm eff}$ and $L$ parameters may be estimated from observational spectra \citep[e.g.,][]{Saumon2008L-T-dwarfs,cushing2005infrared}. In addition, objects with masses 20--80~$M_\mathrm{J}$ and age$\geqslant$0.2~Gyr have radii within a range of 0.7--1.3~$R_\mathrm{J}$ \citep[e.g., see figure 3 by][]{Burrows2001review}. As a result, age, gravity, and mass are the three puzzling parameters as younger, lower mass objects can have comparable luminosity and spectra to older, higher mass objects.


There have been attempts to disentangle this $M-t-g$ degeneracy by employing atomic and molecular spectral features as a probe. For instance, \citet{McGovern2004} have proposed to estimate the gravity and age by comparing the absorption features of Na{\tt I}, K{\tt I}, Cs{\tt I}, Rb{\tt I}, TiO, VO, CaH, and FeH at $\sim 0.7-1.25$~$\micron$. Following that, \citet{Kirkpatrick2008-LiTest-Ldwarfs} used the strength and the presence-or-absence of gravity-sensitive features of Na{\tt I},  K{\tt I},  CaH,  TiO/VO ratio, and H$_{\rm \alpha}$ emission to determine the youth and gravity of a large sample of late-M through L dwarfs. \citet{Allers2013} also used FeH, VO, K{\tt I}, and H-band continuum shape to probe the gravity of several ultracool M5--L7 dwarfs.

\section{Modeling M-L-T dwarf Atmospheres: Cloudy and clear}
\label{sec:modelingflux}

In this work we aim to explore the utility of the Li-bearing species to help disentangle age, mass, and gravity in brown dwarfs by providing a well defined mass boundary.  Figure~\ref{fig:g_T_t_M} highlights which Li species would denote which regions and where the absence of Li would signify a high mass. Details on the evolutionary and atmospheric models are provided in \S\ref{sec:ModelingDetails}, following by discussion on the results in \S\ref{sec:flux} and \ref{sec:atomic-Li-line}.

\subsection{Modeling Details} \label{sec:ModelingDetails}
Here we utilize the Sonora-Bobcat generation of atmosphere models computed for brown dwarfs with clear atmospheres by \citet[][]{Marley2021}. These are supplemented by a few additional cloudy models computed specifically for this study, following the same methodology and the cloud treatment  described in \citep{Ackerman2001clouds}. For each model of interest we compute a high resolution spectrum using {\tt PICASO}, an open source radiative transfer code \citep{Batalha2019picaso, Batalha2019zndo-picaso}\footnote{\href{https://natashabatalha.github.io/picaso/}{https://natashabatalha.github.io/picaso/}}. 

We used our generated opacities for Li (g), LiH, LiCl, and LiF as well as  opacities from \citep{Freedman2014} \citep[see Table 2 in][for details]{Lupu2014} for other species mentioned in \S\ref{sec:LiChemistry}. We did not investigate LiOH due to the lack of a line lists, although theoretical {\it ab intio} studies suggest that the LiOH main rotation-vibration bands are located at 2.6, 10.8, 31.3~$\micron$ \citep[e.g.,][]{Bunker1989LiOH,Koput2013LiOH}. The evolution calculation is from \citet[][]{Marley2021}.

\subsection{Impact of L{i}, LiCl, LiF, LiH on Brown-dwarf Spectra}\label{sec:flux}

\begin{figure*}[htb!]
\centering
\includegraphics[scale=.7]{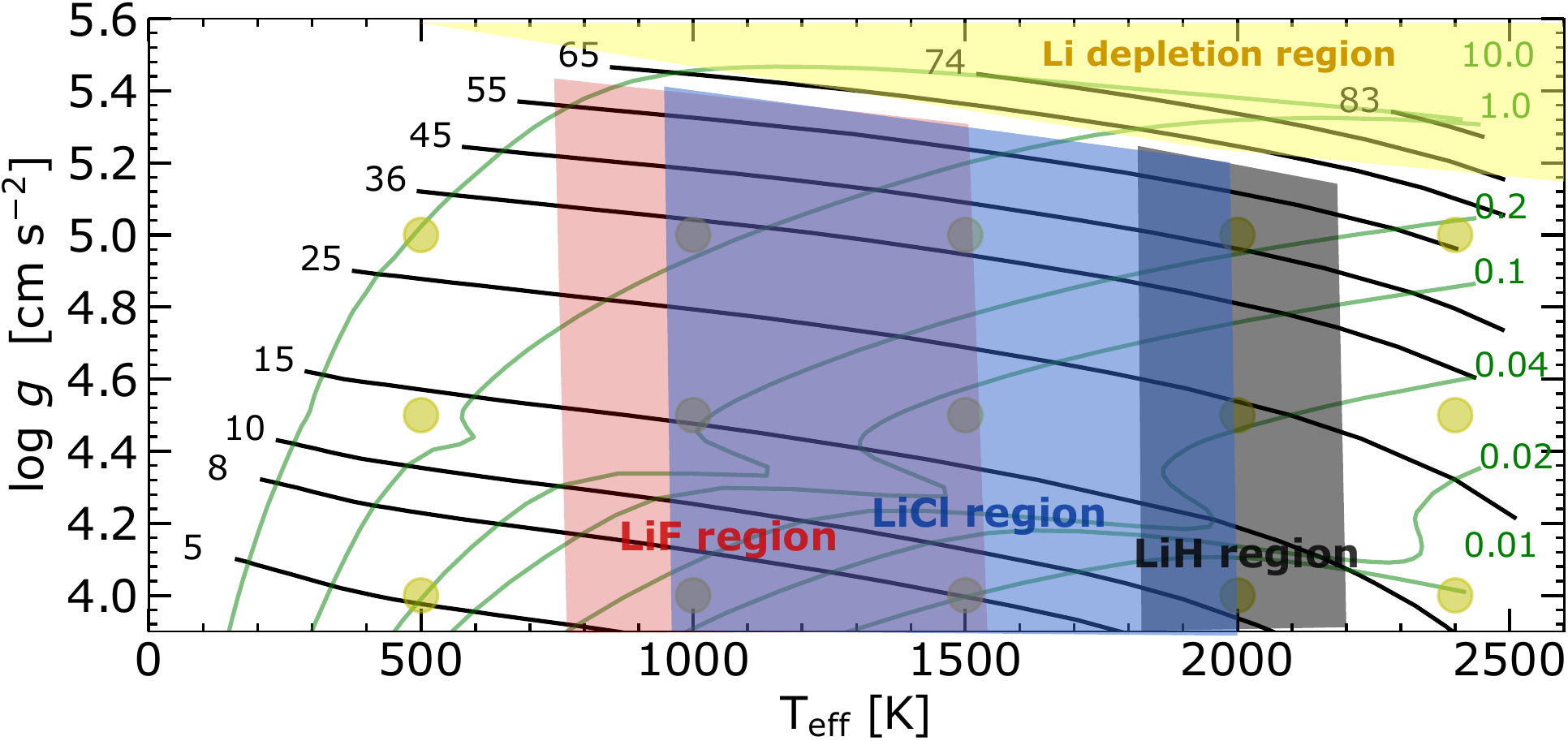} 
	  \caption{ Evolution tracks (black, moving right to left with increasing age) showing gravity as a function of effective temperature for
	   substellar-mass objects. Masses (black, in $M_\mathrm{J}$) and ages (green, in Gyr) are labeled. Yellow circles denote individual
	   radiative-convective models shown in the other figures
	   ($T_{\rm eff}=$ 500, 1000, 1500, 2000, $2400\,\rm K$, and  surface gravity of  $\log g= 4.0,  4.5\,\rm{and}\, 5.0$  for cloudless models. The Sonora-Bobcat evolution grid models are adopted from \citet{Marley2021}. Objects in the yellow area are massive and old enough to burn protosolar Li atoms. LiH, LiCl, and LiF spectral features become apparent in the black, blue, and red areas, respectively. For the strength of their features, see Figs.~\ref{fig:Flux_ON_OFF_cloudModel_g5}--\ref{fig:Flux_ON_OFF_cloudfreeModel_g4.5-5}.}
	  \label{fig:g_T_t_M} 
\end{figure*}

We investigate the influence of Li species on the emergent spectra of candidate brown dwarfs with  protosolar abundances. We consider effective temperatures ($T_{\rm eff}$) of 500, 1000, 1500, 2000, and 2400~K and $\log g= 4.0, 4.5,$ and 5 to span the range of conditions of interest where Li species may be detected. The cloudy models are computed for this study for all the aforementioned $T_{\rm eff}$ but with $f_{\mathrm{sed}}=3$ and $\log g=5$. The sedimentation parameter, $f_{\mathrm{sed}}$, controls the particle size and vertical extent of the cloud layer \citep{Ackerman2001clouds}. Aluminium Oxide (\ce{Al2O3}), forsterite (\ce{Mg2SiO4}(s)), and Fe(l,s)) cloud species are included. The location of the modeled objects on the evolutionary model space is indicated with yellow circles in figure~\ref{fig:g_T_t_M}.

\begin{figure*}[htb!]
\centering
\includegraphics[scale=.7]{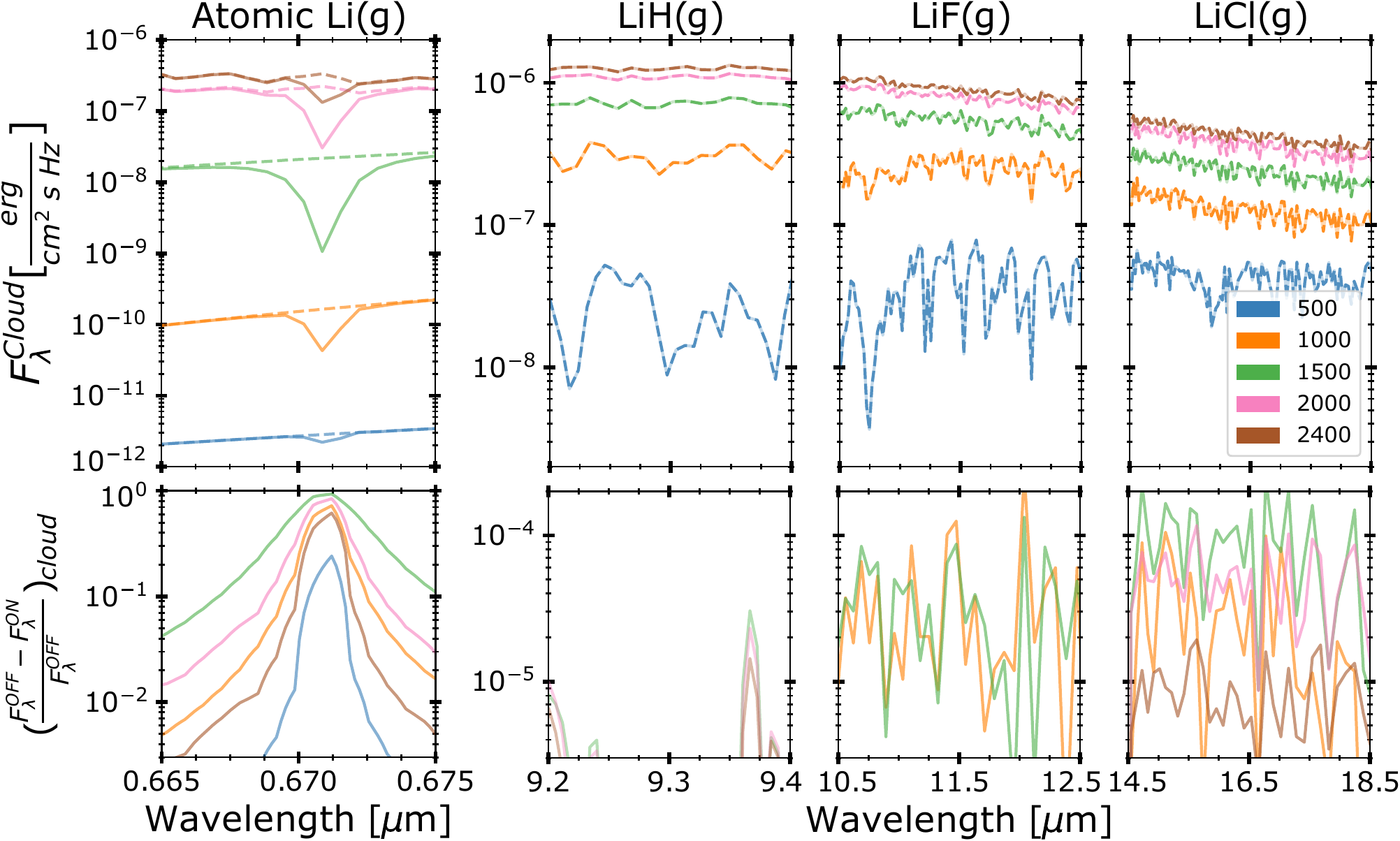} 
	  \caption{Overview of  model spectra as computed for cloudy models with  various $T_{\rm eff} $ in 500-2400$\,\rm K$ range with $\log g=5.0$ and $f_{\rm sed}=3$. 
	  The top row of panels shows the computed spectra as computed with ($F^{\rm ON}$, solid) and without ($F^{\rm OFF}$, dashed)  Li, LiH, LiF, and LiCl. Note the difference in the vertical axis between Li and the other species; in contrast to Li-bearing molecules, atomic Li shows an easily apparent difference between its $F^{\rm ON}$ and $F^{\rm OFF}$ fluxes.
	  The lower row of panels shows the relative flux ratio, which is the difference between $F^{\rm ON}$ and $F^{\rm OFF}$ normalized by $F^{\rm OFF}$  ($\mathcal{F}_{\rm R} \equiv \frac{F^{\rm OFF}-F^{\rm ON}}{F^{\rm OFF}}$).
	  All spectra are smoothed for clarity. LiF and LiCl could be an indicator of dwarfs with $M < 65 M_{\rm J}$ at $T_{\mathrm{eff}}$ ranges of 1000--1500~K and 1000--2000~K, respectively. Their absence in the observational spectra would be an indicator of higher mass.}
	  \label{fig:Flux_ON_OFF_cloudModel_g5} 
\end{figure*}

\begin{figure*}[htb!]
\centering
\includegraphics[scale=.6]{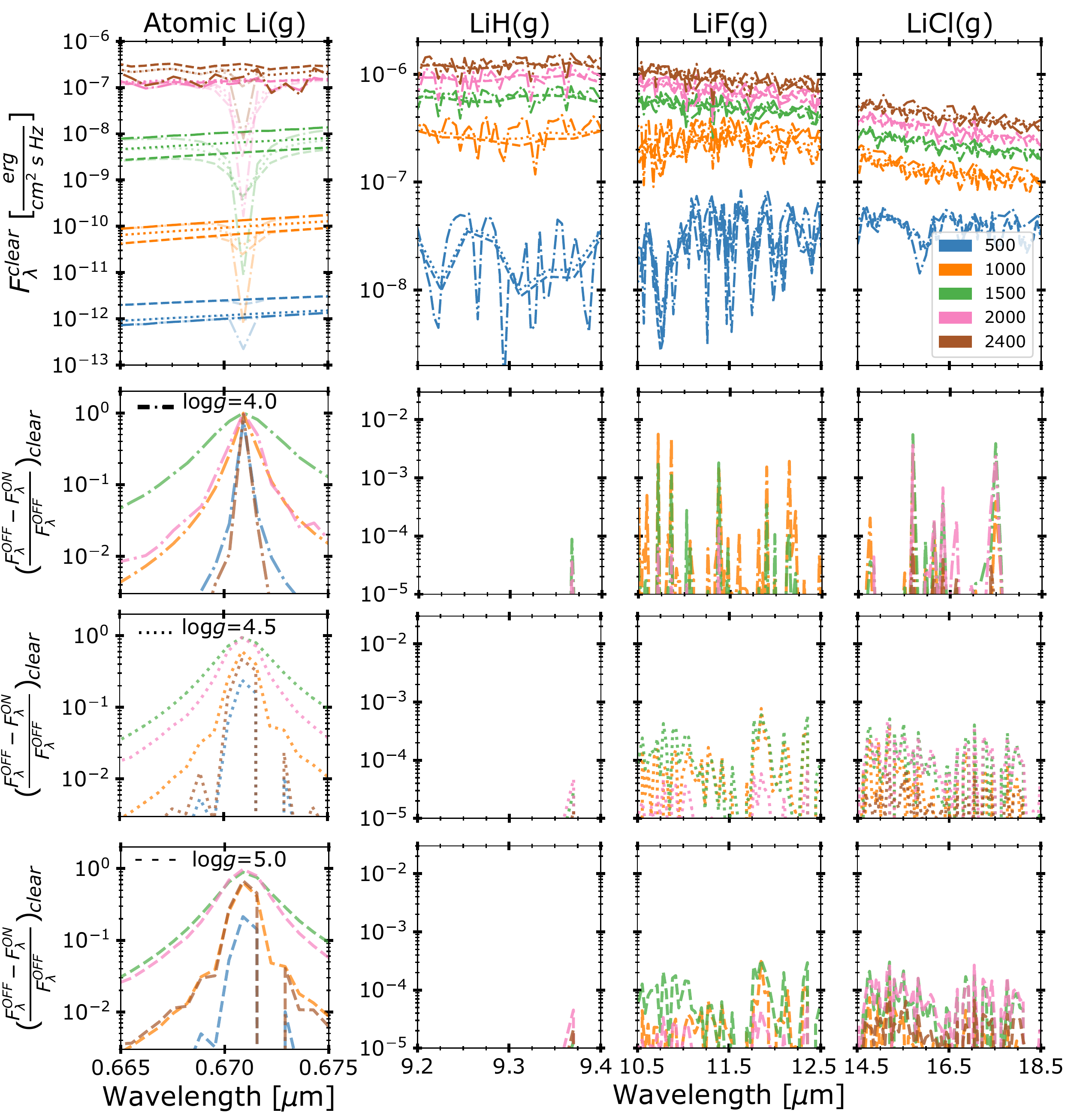} 
	  \caption{
	  Similar to Figure 5, but for cloudless spectra and  different $T_{\rm eff}$ in 500--2400~K range with $\log g$=4.0 (dashed-dotted), $\log g$=4.5 (dotted) and 5.0 (dashed). 
	  (Top panels) The presence ($F^{\rm ON}$, solid) and absence ($F^{\rm OFF}$, dashed) of Li, LiH, LiF, and LiCl. As with the cloudy model cases, atomic Li shows a noticeable difference between its $F^{\rm ON}$ and $F^{\rm OFF}$ spectra. The bottom three rows show
	  the relative flux ratio, which is the difference between $F^{\rm ON}$ and $F^{\rm OFF}$ normalized by $F_{\rm OFF}$ ($\mathcal{F}_{\rm R} \equiv \frac{F^{\rm OFF}-F^{\rm ON}}{F^{\rm OFF}}$) for $\log g$=4.0, 4.5 and 5.
	  All spectra are smoothed for clarity. The spectral features of LiF and LiCl are stronger at lower surface gravity. LiH is not sensitive to gravity and its $\mathcal{F}_{\rm R}$ is lower than other Li species.
	  LiF and LiCl are most detectable in dwarfs with $M < 65 M_{\rm J}$ at $T_{\mathrm{eff}}$ ranges of 1000--1500~K and 1000--2000~K, respectively. Their absence in the observational spectra would be an indicator of higher mass.
	  }\label{fig:Flux_ON_OFF_cloudfreeModel_g4.5-5} 
\end{figure*}

The presence, absence, and strength of the Li-species spectral features on emergent spectra are assessed for different effective temperatures and surface gravities to see if they can productively be used as a high mass indicator in brown dwarfs. For each case, we perform the model twice by in turn including (ON) and excluding (OFF)  each molecule or atom in our calculation of the emergent thermal spectrum.  This was done to isolate the effect of each species. The results for cloudy and cloudless models are illustrated in Figures~\ref{fig:Flux_ON_OFF_cloudModel_g5}~(for $\log g$=5) and  \ref{fig:Flux_ON_OFF_cloudfreeModel_g4.5-5}  (for $\log g$=4.0, 4.5 and 5). In Figure \ref{fig:Flux_ON_OFF_cloudModel_g5}, the top panel represents the 500--2400~K synthetic thermal fluxes with Li, LiH, LiF, and LiCl ON (solid) and OFF (dashed). The bottom panel shows the relative flux ratio, which is the difference between $F^{\rm ON}$ and $F^{\rm OFF}$ normalized by $F^{\rm OFF}$ (hereafter, $\mathcal{F}_{\rm R} \equiv \frac{F^{\rm OFF}-F^{\rm ON}}{F^{\rm OFF}}$). In Figure \ref{fig:Flux_ON_OFF_cloudfreeModel_g4.5-5}, the $\log g$=4.0, 4.5 and 5 are shown with dashed-dotted, dotted, and dashed lines, respectively, which includes both ON and OFF cases.

Our focus is to ascertain  whether the presence or absence of each Li species under different atmospheric effective temperatures and gravities can be detected in the synthetic emergent spectra. Below we discuss each species in turn in order to better interpret these spectra and how they relate to the object's mass.

Atomic Li is the predominant form of Li species at high temperatures. The Li resonance line is shown in the left panel at 0.667--0.675~$\micron$. Since Li has a large opacity (see Figure~\ref{fig:LiX-XS}), its absorption is predicted even at  $T_{\rm eff}<1000$~K in both cloudy and cloudless models. Given the fact that Li absorption is narrow and in a faint part of the spectrum, it requires a resolution of $R \geqslant 1200$ and signal-to-noise (S/N) of $>50$ to be detected \citep{Kirkpatrick2008-LiTest-Ldwarfs,Martin1999Litest-BD}. In particular, for T dwarfs, large telescopes and a significant amount of exposure time are required.

In the cloudy model (Fig.~\ref{fig:Flux_ON_OFF_cloudModel_g5}), the Li $\mathcal{F}_{\rm R}^{Li}$ ratio has the largest value which is $\sim 100\%$ at the line center for  $T_{\rm eff}$=1500~K. In contrast, this ratio decreases to $\sim~20\%$ for $T_{\rm eff}$=500~K, and this might be why Li has not been observed in late-type T dwarfs (see \S\ref{sec:atomic-Li-line} for more details). Note that this Li $\mathcal{F}_{\rm R}$ ratio is about 3--4 order-of-magnitude larger than LiH, LiF, and LiCl, which shows the significant difference between presence and absence of atomic Li in the model spectra. 

In the clear model (Fig.~\ref{fig:Flux_ON_OFF_cloudfreeModel_g4.5-5}), the Li $\mathcal{F}_{\rm R}$ ratio for $T_{\rm eff}$=1500--2000~K is $100\%$ for all surface gravities  ($\log g$=4.0, 4.5, 5) at the line center. The behavior of the $\mathcal{F}_{\rm R}^{\rm Li}$ value with the wavelength in the wings (i.e., $d\mathcal{F}_{\rm R}^{\rm Li}/d\lambda$) could also be an indicator of the gravity because $d\mathcal{F}_{\rm R}^{\rm Li}/d\lambda$ is larger for $T_{\rm eff} =2400$ and 500~K in models with smaller surface gravity. Comparing the $\log g$ of 4.0, 4.5 and 5 in Figure \ref{fig:Flux_ON_OFF_cloudfreeModel_g4.5-5} shows the negligible impact that surface gravity has on the Li line core. This change in the wings with surface gravity is appeared in the figure.

Comparing cloudy and clear models for Li shows some remarkable change in the Li $F^{\rm OFF}$ and $F^{\rm ON}$ fluxes, and slightly in $\mathcal{F}_{\rm R}^{\rm Li}$ ratio. Clouds have the largest impact on atmospheres with $T_{\rm eff}$ in 1000--2000~K, although the cloudy models still exhibit the atomic Li line.

LiH is the molecular form of Li at high temperatures and its spectral feature is located at 7--10~$\micron$, but its $\mathcal{F}_{\rm R}^{\rm LiH}$ ratio is greatest at $\sim$9.36~$\micron$ according to our results. 
The $\mathcal{F}_{\rm R}$ ratio at $T_{\rm eff} >$1000~K is $\sim 0.001-0.005\%$ which would be challenging to detect. In the cloudy model, the highest $\mathcal{F}_{\rm R}^{\rm LiH}$ ratio is for $T_{\rm eff}$=1500~K, while this changes to 2000~K for the clear model. Assessing both cloud and cloudless models and different surface gravities suggests that LiH is not sensitive to gravity and so could not be employed as a gravity indicator. However,  LiH region (black area) in Figure~\ref{fig:g_T_t_M} shows that its detection can be an indicator of $M<65M_{\rm J}$ and temperature of $\sim$1700--2500~K. Moreover, LiH is present in young objects with $t<1.0$~Gyr.

The LiF absorption feature is most prominent in the spectral range of 10.5--12.5 $\micron$. As it is predicted by thermochemical equilibrium, LiF mostly forms at $\sim$800--1200~K, and its $\mathcal{F}_{\rm R}^{\rm LiF}$ ratio is almost $\sim 0.005-0.5\%$ for T--L type dwarfs with $T_{\rm eff}$ of 900--1600~K. 
According to the cloudy and clear models, LiF can only be detected in atmospheres with $T_{\rm eff}$ in 1000--1500~K. In clear model, $\log g$=4.0 has the greatest $\mathcal{F}_{\rm R}^{\rm LiF}$. According to the evolutionary $T_{\rm eff}-g-t-M$ parameter space (Fig.~\ref{fig:g_T_t_M}), the detection of LiF is a signature of an object with $M\leq65M_{\rm J}$ and $t \leq 10$~Gyr. Lower surface gravity in this space leads to a higher $\mathcal{F}_{\rm R}^{\rm LiF}$ ratio, and hence lower mass and younger age.

The predominant LiCl feature is located at 14.5--18.5~$\micron$. This molecule is present for $T_{\rm eff}$=1000--2100~K, and its  $\mathcal{F}_{\rm R}^{\rm LiCl}$ ratio is $\sim 0.005-0.5\%$. The feature is most detectable at $T_{\rm eff}$=1500--2100~K. Both cloudy and clear models show that the LiCl feature and its $\mathcal{F}_{\rm R}^{\rm LiCl}$ ratio are sensitive to surface gravity.  $\mathcal{F}_{\rm R}^{\rm LiCl}$ is strongest for the lowest gravity case considered,  $\log g = 4.0$. Such objects would be young and low mass (i.e, $M\leq15M_{\rm J}$ and $t \leq 0.1$~Gyr).  LiCl is less detectable in higher gravity objects. Figure~\ref{fig:g_T_t_M} illustrates the relation between the detection of these species with mass, gravity, and age.  Note that both LiCl and LiF features are  present in the range $T_{\rm eff}=\sim900-1600~K$, and their $\mathcal{F}_{\rm R}$ ratio is higher for objects with lower surface gravity.

\citet{Weck2004} investigated the signature of LiCl on dwarfs with $T_{\rm eff}$=900--1500~K and $\log g$=3.0, 4.0, and 5.0 using the PHOENIX code with solar abundance.
They concluded that the strongest $\mathcal{F}_{\rm R}$ LiCl features are expected in T dwarfs with $T_{\rm eff}$=1200~K and $\log g$=3.0. According to the evolutionary models, only very young objects with age$<$0.01~Gyr and mass$<10~M_{\rm J}$ could have $T_{\rm eff}$=1200~K and $\log g =3.0$. In comparison to their work, our main focus here is the study all Li species but for older objects with greater surface gravity and so with greater mass (i.e., $5M_{\rm J}\leq M \leq65M_{\rm J}$). We also assess the impact of clouds on the detection of the various Li species using the latest and most accurate their opacity data. 
 
In another study, \citet{Pavlenko2007Li}   calculated the atomic Li resonance abundance in LP~944-20 M-dwarf atmosphere using VLT/UVES and AAT/SPIRAL observed spectra. They used DUSTY and COND model atmospheres from \citet{Allard2001} with solar metallicity, and included Li, LiOH, LiH,
LiF, LiBr, and LiCl species, and obtained $\log N( {\rm Li})=3.25 \pm 0.25$. They found that atomic Li is predominant Li-bearing species in objects with $T_{\rm eff}$=1800--2400~K with $\log g=4.5$.

\subsection{Comments on Atomic Li line in T dwarfs}\label{sec:atomic-Li-line}

\begin{figure*}[hbt]
\centering
\includegraphics[scale=.7]{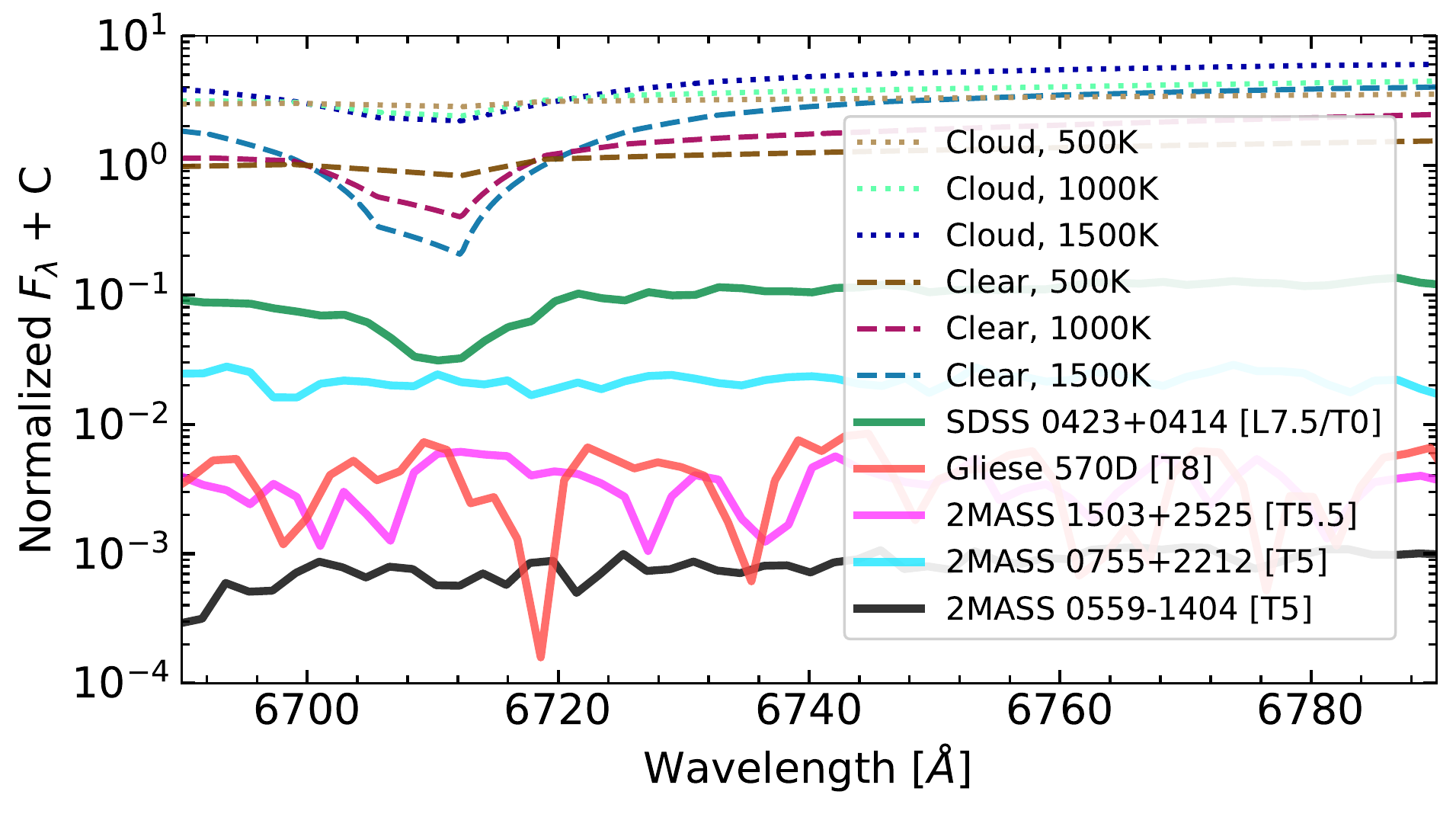} 
	  \caption{Comparing a set of observed T dwarf fluxes with synthetic spectra from Sonora cloudy and cloud-free models ($\log g$=5). Theoretical evolution models predict the detection of atomic Li 6708 $\AA$, whereas many T-dwarf spectra have not shown any Li signature. For example, observed fluxes for Gliese 570D, 2MASS 0559-1404, 2MASS 0755+2212, 2MASS 1503+2525 do not show any Li absorption.
	  Note, the Near-Infrared part of the spectrum is T0, while the optical part of the SDSS 0423-0414 (SDSS~J042348.57-041403.5) flux is classified as L7.5 and so it shows the Li line with EW=11~\AA  \citep{Kirkpatrick2008-LiTest-Ldwarfs,Vrba2004} { (spectra are from \citet{Burgasser2003Tdwarfs} and \citet{Kirkpatrick2008-LiTest-Ldwarfs})}.}
	  \label{fig:Flux_Li_Burgasser} 
\end{figure*}

Figure~\ref{fig:Flux_Li_Burgasser} presents a set of observed emergent spectra for different T dwarfs, encompassing Gliese~570D (T8), 2MASS~1503+2525 (T5.5)  \citep{Burgasser2003_2MASS-Tdwarfs}, and 2MASS~0755+2212 (T5) \citep{Burgasser2002Tdwarfs}. These T dwarfs apparently do not show any atomic Li absorption in their spectra. In this figure, SDSS~0423-0414 (SDSS~J042348.57-041403.5) is obtained from \citet[][]{Kirkpatrick2008-LiTest-Ldwarfs}, and is the only object in these spectra that shows Li absorption { with a pseudo equivalent width measurement of 11~\AA}. Note, this object is a peculiar one because its optical part is classified as L7.5 while the Near-Infrared part is T0 \citep{Vrba2004}.  We also showed our synthetic spectra from the clear (dotted) and cloudy (dashed) models to highlight the presence of atomic Li from our modeling work.

The key difference between the observed and modeled spectra is  the atomic Li line at 6708 \AA. At low temperatures, atmospheric thermodynamic simulations expect Li to predominantly be found in the atomic form only at high $P$ \citep{Lodders1999}  (Figure~\ref{fig:VMR}). Nevertheless our grids of atmospheric cloudy and clear models predict the presence of the atomic Li line even at $T_{\rm eff} \simeq$600~K \citep[e.g.,][]{Allard2001,Burrows2002-Tdwarfs}. However, the observed T-dwarf spectra do not show any sign of Li detection at $T_{\rm eff}\leqslant$900~K \citep[e.g.,][]{Kirkpatrick2000, Burgasser2003Tdwarfs}. { Theoretically this cannot simply be due to Lithium burning having removed Li as objects massive enough to burn Li are not expected to cool below 900 K in the age of the universe \citep[e.g.,][]{Marley2021}, although dynamical masses for at least some T dwarfs come close \citep[e.g.,][]{Sahlmann2020,Brandt2020}}.

Other possible explanations for the predicted atomic Li not being found in the data include incorrect Li thermochemistry data and  clouds obscuring the deeper, higher pressure regions where atomic Li resides. The accuracy of thermochemical data for key reactions such as LiOH+H =Li+\ce{H2O} and LiH+H=Li, as well as the impact of condensation on the abundance of Li gas needs further investigation. Greater cloud opacity in the optical than considered here might also obscure the Li line.  Additionally, since this Li line is located in the faint part of the optical spectrum, higher S/N ratio spectra should be acquired to more robustly test if the feature is indeed missing. The detection of Li at 670.8~nm at $T_{\rm eff}<$1000~K deserves further observational and theoretical attention.

\section{conclusion and Future Work}\label{sec:Conclusion}

Since the discovery of the first brown dwarfs, numerous massive cool dwarfs have been detected. However ascertaining precise masses have been a challenge to the degeneracy between mass, gravity, and age for these objects. Therefore, the calculated mass of many ultracool dwarfs including  Gl~229B and Gl~510D is still under debate and spans a wide range. 

Traditionally, the Li test has been used to distinguish between low mass stars and brown dwarfs with similar
spectral types. At lower temperatures such as L-T dwarfs,  LiF and LiCl molecules form. Because LiF and LiCl
features are located in regions where brown dwarfs are brighter than 670~nm, they could in principle be used
as mass indicators at these lower effective temperatures.

In this article, we investigated the spectral signatures of LiCl, LiF, and LiH as well as  atomic Li using cloudy and cloudless atmospheric models. We showed that LiCl is present at an effective temperature of $\sim$1000--2000~K in cloudy atmospheres and $\sim$1500--2000~K in clear atmospheres. Hence, the LiCl spectral features should be found in L-T type dwarfs with masses below 65~$M_{\rm J}$. The lack of LiCl at these $T_{\rm eff}$s would be an indicator of larger masses with concomitant older ages and larger surface gravities. The LiF feature at ~10.5--12.5 $\micron$ is another indicator of masses lower than 65~$M_{\rm J}$ and  $T_{\rm eff} \leqslant 1600~K$.
LiH is a stable molecule at high temperatures (e.g., early-L--M dwarfs), however its signature is confined to a narrow spectral interval of $9.35-9.4~\micron$. We could not assess the contribution of LiOH because of the lack of appropriate opacity data. 
In ultracool T dwarfs such as Gl~229B with a large mass uncertainty, the detection of LiF and LiCl would confirm the object has mass is lower than 68~$M_{\mathrm{J}}$.

The relative flux ratio, $\mathcal{F}_{\rm R}$, of LiF and LiCl is very small, thus the detection of these species requires  high SNR spectra in the mid-IR. However, their strongest spectral bands are found in the relative bright region of the infrared thermal  with less overlap with other spectral features than the atomic Li line in the optical, making them potentially favorable mass indicators. The $JWST$ MIRI instrument covers the wavelength range of $5-28 \micron$, and its spectrograph will enable medium-resolution spectroscopy with $\lambda/\Delta\lambda \sim 2000-3700$, and so MIRI might be capable of detecting these Li features.    

The high-resolution cross-correlation technique could be another method to detect LiH, LiF, and LiCl molecules. In this method, the number of detected lines is crucial to increasing the signal-to-noise ratio, and so a large number of transitions of these molecules would be important. However,  current ground-based telescopes such as $VLT$ CRIRES+ \citep{Follert2014}  only reach up to 5 micron. Atomic Li lines at 600--900~nm could be detected using $CAHA3.5$ CARMENES (520--1710~nm, R$>$80,000) \citep{Quirrenbach2010}, and the $Subaru$/High Dispersion Spectrograph (298--1016~nm, R$>$90,000) \citep{Noguchi2002}.

Another potential avenue for fundamental research is to improve the thermodynamic data in order to understand the reason for the apparent non-detection of the Li 670.8~nm line in T-dwarf thermal fluxes. Knowing the kinetic data between all Li-bearing reactions as well as their condensation would be another topic to advance for future work.


\section{Supplementary data}
The pre-generated absorption cross-sections (ACS) data from this study will be uploaded in ZENODO upon acceptance.


\section{Acknowledgements}

We would like to thank Dr. Adam Burgasser for his assistance regarding the observational T-dwarf spectra. 
EGN acknowledge Research Computing at Arizona State University for providing HPC resources that have contributed to the research results reported within this paper. 
EGN acknowledge support from HST-AR-15796 to generate Li-species opacities. 
We also acknowledge the ExoMol team for their continued production of large line lists and pertinent data critical to high-temperature atmospheric modeling. 
EGN’s research was supported by an appointment to the NASA Postdoctoral Program at the NASA Ames Research Center, administered by Universities Space Research Association under contract with NASA. 

\section{Software}
We used the publicly available $ExoCross$ code \citep{Yurchenko2018ExoCross} developed by the ExoMol group to generate the cross sections. To perform the radiative-transfer modeling of brown dwarfs, {\tt PICASO} tool was used \citep{Batalha2019picaso,Batalha2019zndo-picaso}. Brown-dwarf thermal structures and cloud compositions were used from  {\tt Sonora} cloudless and cloudy models \citep{Marley2018CloudFree, Marley2021}.

\bibliography{reference_8}

\end{document}